# Photomultipliers as High Rate Radiation-Resistant In-Situ Sensors in Future Experiments SNOWMASS Calorimetry

**David R Winn - Fairfield University***
**Yasar Onel -University of Iowa'**

***SUMMARY:*** In the Energy Frontier we suggest developing high rate (≥100 MHz) finely segmented forward calorimetry preradiators with time resolution <50 ps which will survive the first ~1-2 λ of incident high radiation doses, protecting forward calorimeters (3<η<6; less than 5° to the beam) behind them from radiation damage, with high granularity, high rate capability and ~40ps time resolution (4D-calorimetry) providing lepton/photon ID and measurement. In the Intensity Frontier beam particle selection, such as tagged neutrino and kaon beams, and lepton violation experiments with muons require very high rates. Cosmic Frontiers requiring low power, non-cooled calorimetry or optical detection that can keep track of particles or photons arriving at 100's of MHz, and survivable for years in space radiation may also benefit. The basic research is to use compact channelized PMTs with quartz or other radiation resistant windows with metal envelopes as an in-situ sensor, directly coupled to Cerenkov (or radiation-resistant scintillator) tiles, utilizing the dynode signals as a potentially compensating 2nd signal, and with no active electronics. If successful, directions include proposals for high SE yield mesh dynode activator materials such as GaP or B-doped diamond films with 25 SEe at 300 eV electron energies, and possibly for compact low cost tile SE sensors - no photocathode - far easier to fabricate than PMTs: all metal final assembly in air, brazed seals; bake-out 900°C; pump out with tip-off - vacuum 100x higher than PMTs. Such sensors have many applications beyond HEP, in research, medicine, industry and defense.

***Introduction:*** We propose compact PMT systems with rate capability >100MHz, able to survive ~GigaRad radiation doses, and time resolution <50 ps as in-situ calorimeter sensors. This study investigates high gain quartz window metal envelope compact multi-anode photomultipliers as direct in-situ light and particle sensors. In addition to the Cerenkov light generated in the quartz window or coupled to Cerenkov or scintillator tiles, the photomultiplier dynodes in hadron showers generate signals from ionizing particles traversing the PMT dynodes. The dynodes have small sensitivity to mips(minimum ionizing particles), but low βγ low energy ion fragments, α, p±, π±, x-rays, neutron-knock-on particles penetrating the PMT cause secondary emission electrons (SEe) at higher efficiency than mips due to higher dE/dx in the dynodes, forming a possible dual readout correction to an energy signal as combined with Cerenkov signal. Such PMT systems could be used in-situ as detectors in pre-radiators, muon track ID or in-situ calorimeter sensors with transverse spatial resolution as small as 3x3mm for energy flow reconstruction (4-D calorimetry) which survive high radiation damage in the forward regions (3<η<6) of collider experiments, where solenoidal magnetic fields are minimal and able to be shielded to 1T or less, and where the occupancy of 1cm² of detector is 100% per collision. Future colliders with 10 ns/100 MHz beam crossings are planned, and sensors must operate at those rates.

***Energy Frontier:***

• Calorimeters in the Forward Region (7≥η≥3) at near and far future pp, ep, e+e-, e-ion, and muon colliders will need to measure high-energy depositions at irradiation levels approaching/exceeding GigaRads and >$10^{17}$ neutrons/cm² in the front 1-2 λ of calorimeters, especially in the very forward regions. In planned LHC energy and luminosity upgrades over 20 years, at 3,000-5,000 fb⁻¹, the 1 MeV equivalent neutron flux is ~$10^8$/cm²/s at η≥4.9 and at ~10 m from the IP [1] [2] (see Fig. 1).

* Corresponding Author: winn@fairfield.edu ORCID 0000-0003-2637-5743

• At 11 m from the interaction, the η≥3 forward region at HLHC/SLHC, *the occupancy of a 1 cm² patch of calorimeter is 100% per crossing,* with pileups of ~150-900 events/crossing at highest luminosities.
• In future colliders, the calorimeter channels must operate at 40 MHz (present LHC), *with some proposed colliders or upgrades up to 100 MHz,* with calorimeter hysteresis pulse-pulse <2%.
• For pileup mitigation, a time precision for mips must be ~30-40 ps or better, with energy flow capability.
• Energy scale ~0.1 GeV (1/4 minimum ionizing particles - MIPs) to ~10 TeV, a dynamic range of ~$10^5$.

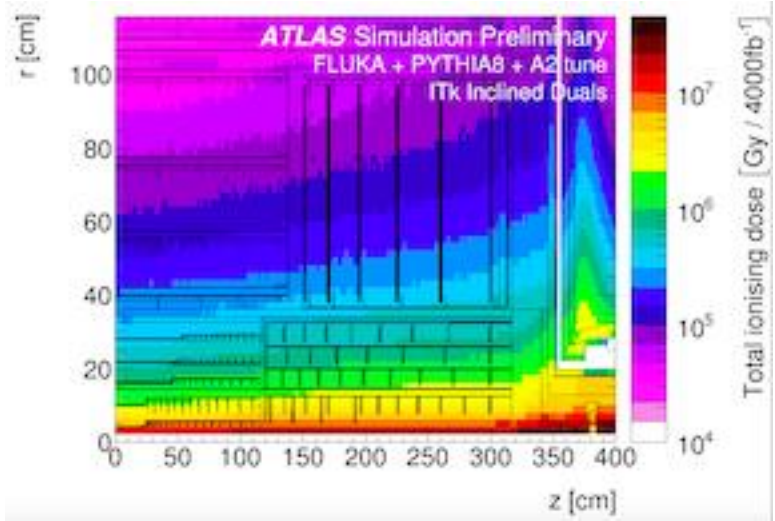

***Fig. 1:*** *ATLAS simulation 4,000 fb⁻¹[3]. At η≥ 4.9, a calorimeter dose ≥ GigaRads in the 1ˢᵗ ~2λ.*

Studies of vector-boson scattering and fusion require very forward jet tags and precision measurement of the subsequent vector boson decays. The η reach necessary to obtain full statistics will grow with the log of increasing energy. In the next stages of experiments at LHC with luminosity and energy upgrades, the forward region η≥3 is becoming more important for discovery and measurement of important and rare physics processes. With energy upgrades, a larger detector reach in η is required for confirming QCD predictions and for high statistics. The Forward Region as 2.8<η<5.9 has emerged as favorable for new physics activity, in kinematics of signals, the need for high statistics, and fuller understanding of underlying events and the initial parton distribution functions (PDFs) over the x/Q² space. Forward region |η|>2.9 calorimetry in high energy physics is crucial for many high energy processes. The physics includes:

- *VBS-Vector Boson Scattering* and *VBF-Vector Boson Fusion:* $W_L W_L$, $W_L Z_L$, and $Z_t Z_L$ interactions directly probe EWSB, dominated by the transverse gauge-boson components of the W and Z. It is essential to show that the s-wave scattering amplitude of WW, WZ, ZZ is damped by the Higgs, protecting unitarity. Excesses in the longitudinally polarized channel are indicative of new interactions in EWSB. Because color is not exchanged, the non-interacting quarks produce a distinctive signal of relatively low central region activity with 2 very forward jets from the fragmenting protons, with jet rapidity separations Δη_jj > 4.5-5, that helps disentangle signals from more centrally produced multijet QCD backgrounds.[4]

- *Higgs:* Vector Boson-Boson interaction can fuse into a single Higgs or di-Higgs. Higgs' produced in this process lead to ~15% of the leptons from Higgs decays have one of the leptons within the tagging jet cone. Measurement requires in excess of ~1000 fb⁻¹. Operation of a forward calorimeter will require better jet definition, such as could be provided by a front end added to or replacing the existing forward calorimeters in LHC upgrades. Forward jet tags enable studies of the triple Higgs coupling H−>HH; Higgs to invisible, such a dark matter H−>XX; Higgs decays H−>ττ (eventually μμ). S/N requires jets at η>2.8 (η=3.1), and a rapidity difference between tagging jets Δη_jj > 5 for cleanest S/N. The boosted 125 GeV Higgs at higher √s enhances Higgs decay activity in the forward region.

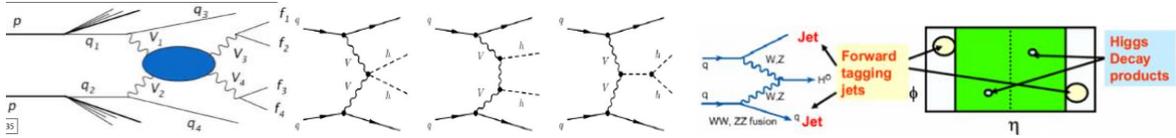

***Figure 2:*** Diagrams for vector boson scattering, Higgsstrahlung, and vector boson fusion into Higgs and Di-Higgs. Cartoon of the forward jets and H decay products in regions of η.

In anticipation of LHC energy and luminosity upgrades, the CMS collaboration, for example, has formed a Task Force under the direction of A.Skuja (Maryland)[5] to study a Phase III upgrade to extend higher precision calorimetry to 2.8≤η≤5.1, and timing to η−>7. The CMS quartz fiber forward calorimeter[6],[7] is crucial for luminosity measurements, forward jets and hermeticity, but unacceptable levels of damage in the region 4≤η≤5 will occur and performance will degrade from 2.9≤η≤4 after 1000 fb⁻¹. The damage can be mitigated/protected by an extra ~1.5 Lint in front of the calorimeter, a cylinder about 40cm in radius which could be designed to be replaced ~every 5 years. At the present time, *few if any* of the basic sensors used in LHC forward calorimetry (collected ions/electrons e.g. Si, LArgon, gasses; optical signals from scintillator or Cerenkov radiators *as detected by SiPMs*) can *either* survive the raddam *or* operate sufficiently fast e.g.−>100 MHz, with large dynamic range and with low hysteresis crossing-to-crossing, *or both*, if operated inside of very forward calorimeters at LHC upgrades in the next decade. For example, SiPM even with active quenching at present cannot operate at 40 MHz with a dynamic range from 0.25 mip (~2 MeV) to 20 GeV without severe hysteresis. In-situ operation is usually a requirement for particle/energy flow, such as the high granularity Si diode calorimeter upgrade in CMS(HGCAL)[8], scintillators and others[9].Unless there is sufficient signal to drive cables to remote electronics, front end electronics required to be intimately proximate to the sensors, must also withstand the fierce radiation dose to avoid both single event upsets and permanent damage. In-situ forward calorimeter sensors at LHC are impractical to fix or replace, becoming so radioactive that humans cannot work on them. Calorimeters must operate at minimum for 10 years with few failures. Detectors that must be cooled, such as Si diodes, SiPM or LAr, require thermal shielding that limits some issues of hermeticity or placement at small radius in the very forward region. Some cooling systems also have issues of radiation induced radioactivity of some circulating coolants. The required dewar thickness limits the smallest inner diameter a calorimeter can be placed radially, for absorbing particles 4.5<η≤6.

To extend the lifetime of forward calorimetry above η>3, requiring high radiation resistance, low power and passive cooling while preserving energy, rate and time resolution, enabling power flow/4D calorimetry, we propose a high radiation resistant calorimeter front compartment, about 1.5-2 λ and 20-25 Xo deep, with ~15%/√E e-m energy and position resolution (3mm) capable of counting at 100MHz. This depth provides sufficient shielding for following calorimeters, and sufficient granularity to tag/ID photons, electrons and muons. We propose studying PMTs as direct in-situ sensors, coupled to quartz or synthetic sapphire tiles (or fast scintillators with sufficient radiation resistance). With multianode metal envelope quartz window MAPMT, cell sizes of ~3x3 mm² are feasible.

***Intensity Frontier:*** High rates, Time-of-flight and radiation resistance are part of many fixed target experiments and factory-collider experiments. Examples include: LHCb, Belle/BES-III or future factory calorimeter, TOF, polarimeter and luminosity systems. Present and future rare K decay and some dark photon (beam-dump) experiments require high rate calorimetry. In the future, tagged neutrino beams to obtain purer beams of electron or muon neutrinos [10], [11], and tagged kaon beams are proposed, requiring >100 MHz counting rates. Muon lepton violation experiments also benefit from very high counting rates.

***Cosmic Frontier:*** About 40 ps precision TOF calorimetry *at low power* could find its place in future high altitude balloon, satellite/space station experiments, where signals rising just above 1 p.e. per 3ns photon backgrounds can benefit. Speculatively, eventually in moon-based experiments for cosmic ray phenomena may need to survive for many years in space-radiation environments. PMTs using active (Cockcroft-Walton or similar) bases require very little power for high gain, and require little or no cooling power.

***PMT, Dynode and Secondary Emission Radiation Resistance:***

*Dynodes and Secondary Emission:*
Metal-oxides as the Secondary Emission(SE) material on the surfaces of metal or ceramic dynodes at present survive ~10 GRad exposures of electron bombardment at the last dynodes. Photomultipliers with quartz or synthetic sapphire windows survived unshielded for decades in space conditions. The LHC SE beam monitors using a thin $Al_2O_3$ film survive $10^{20}$ high energy protons/$cm^2$ with no loss in sensitivity, and can be brought up to air repeatedly also with no change.

Comparative tests on gamma- and neutron- irradiation of industrial photo-multipliers and various PMT window materials confirm that permanent degradation of PMT sensitivity depends almost entirely on PMT window transmission loss [12], and not on photocathode or dynode degradation. Exposures of standard glass PMT to 1- and 2-MeV electrons show large increases in dark current *during the exposure* which are *transient*, shown to be almost entirely due to induced fluorescence in soda-lime, borosilicate or UV glass. Tubes with sapphire or quartz windows but glass tube bodies had substantially lower dark current rise during flood exposure to electrons but returned to normal dark current after the exposure ended [13]. The ATLAS experiment exposed 2 different quartz window-but glass body bialkali PMT to ~3 x $10^{14}$ n/$cm^2$ [14]. The dark current just after exposure increased by a factor of 10 in a 13 mm diameter PMT and by a factor of 7 in a 10 mm diameter PMT, both of which recovered to ~90% in 10 weeks post irradiation. However there was no obvious effect either on the QE of the bialkali photocathode or the gain. Induced light glow from the tube body glass was consistent with the dark current.

*Residual Gas:* A caution is that PMT with residual gas can damage the photocathode by reverse flowing positive ions formed by the dynode electron cloud; these ions directly impinging on the photocathode both cause after-pulses and can damage the few 10's of atomic layers thick photocathode. We will study possible photocathode damage under high exposures using commercial metal envelope PMT with quartz windows.

*Power, Cooling and In-Situ Front End Electronics*: PMTs with gains $\geq$ 2x$10^4$ can drive high impedance (~300 Ohm) PCB strip-lines or balanced twin conductor PCB microstrip lines using both the anode and last dynode signals, so that front end electronics can be removed to the rear or higher radius edges of a very forward or other fast calorimeters, leaving only the PMT in high radiation. Wide-band (10 V/ns slew rate) monolithic unity-gain buffers (voltage follower) removed from the radiation field by high impedance lines can convert the impedance to the typical impedances of front end electronics, linearly preserving the large signal (the PMT is a nearly perfect current source). PMTs using microchannel plates or resistive bases are not considered, since they have RC time constants too large for >100MHz operation and require heat sinking despite superior time resolution.

*Magnetic Field:* The high radiation forward region in most solenoidal colliding beam detectors have a field <1T, where B-aligned channelized dynode PMT proposed operate with at least 10% of full gain, which for many applications are sufficient. As an example, B<0.1T in the CMS solenoid at 11m from the IP. For some in-situ calorimeter PMTs, high permeability ferromagnetic shielding could be incorporated in the absorbers.

***Secondary Emission Sampling Compensation Effects:***

An in-situ calorimeter PMT with a quartz or sapphire window and coupled to a quartz, sapphire, MgF$_2$, or silica aerogel tile may prove to be quasi-compensating, by, in effect, dual sampling shower particles. Cerenkov light is more sensitive to e-m showers ($\beta$–>1). Dynode secondary emission is proportional to dE/dx, peaking for low energy hadron shower particles traversing the dynodes. The SEe yield is a strong function of momentum ($\gamma\beta$), following dE/dx as in the Sternglass formula and peaking for low-$\beta$ particles. The energy corresponding to a MIP (minimum ionizing particle) is near the minimum of SEe yield. The secondary yield (SEY) $\delta$ is given approximately by the Semiempirical Sternglass formula:

$$\delta = f \, \Lambda \, dE/dx \, [1+ (1+5.5E_p/A_p)^{-1}]$$

where $\Lambda$ = e⁻ escape depth; $E_p/A_p$ is particle energy/mass; f is the normalization constant, typically ~0.8 for metal oxides. SEe is dominated by $dE/dx$, increasing with the ratio $A_p/E_p$.

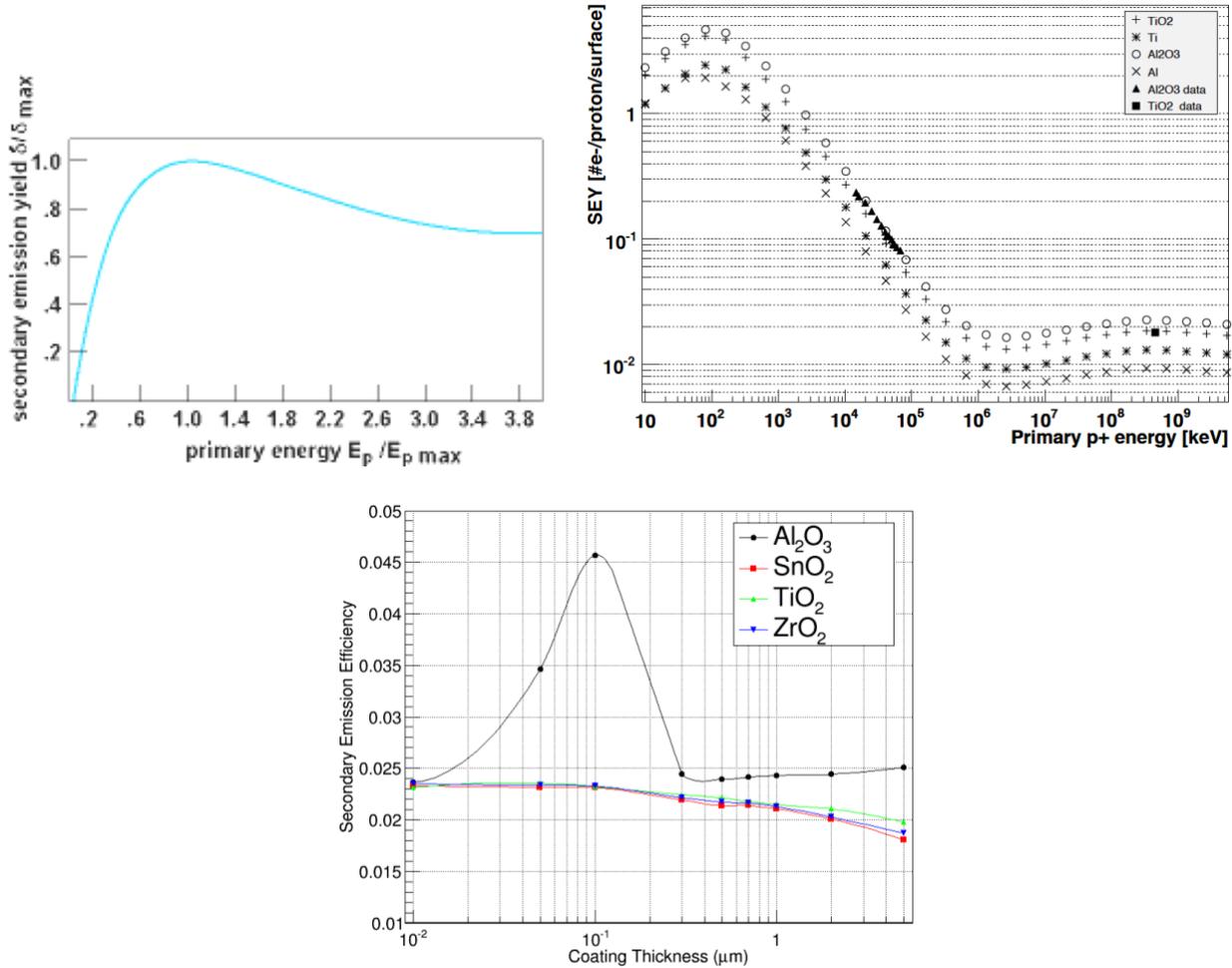

***Figures 3:***

***Top Left:*** The universal curve of SEe emission from metal oxides: the yield is normalized by the maximum yield vs the particle energy normalized to the voltage/energy of the maximum yield emission**;**

***Top Right:*** the secondary yield of protons vs energy. *A a proton mip at 1-2 GeV has 100-200 times lower SEe yield than a ~100 KeV proton. 1-2 MeV π± are near peak SEe yield, whereas 1-2 MeV e± are minimum.*

***Bottom***: **a**verage secondary electron emission efficiency of different secondary emitters (shown in different colors) for MIPs as a function of their thicknesses (right).

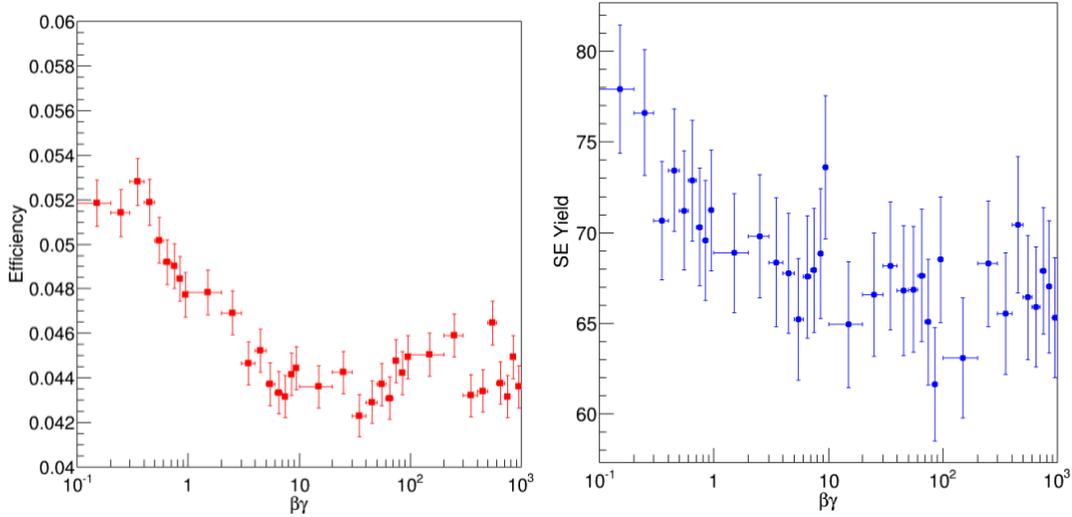

***Figures 4.** Secondary electron emission efficiency (L) and the SEe yield (R) vs βγ of 100-nm Al₂O₃.*

Figures 3 and 4 illustrate the scaled universal SEe yield and the CERN SEe beam monitor data on various metal and metal oxides vs proton kinetic energy. *A 2 MeV proton or a 16 MeV alpha has ~100x-200x more SE yield than a 2 MeV electron, implying that SE surfaces such as dynodes next to hadron absorbers will have more response to low energy ion fragments, protons, pions, or even fast neutron knock-on ions than to electrons.*

Measurements of fission fragments and 4-5 MeV alpha particles impinging on or emerging from heavy metals, metal oxides and photocathodes show SEe yields between 8-20 from 100 nm films [15],[16]. Figure 6 (right) also shows the yield vs film thickness for SE films. Proposals for and tests of SE emission calorimetry are found in [17],[18].

A GEANT4 MC has simulated direct dynode signals, implementing an SE yield process in GEANT4 [19], and confirmed by the PMT data with photocathodes shut off, as described below. *The MC predicts that in a hadron shower, the dynode signal for e/h ~0.8, very different than a pure Cerenkov calorimeter with e/h ~2-3. This difference in response can be used for "dual readout" correction of calorimeter energy.*[20], [21]

Our group has tested standard mesh dynodes in PMT as MIP sensors - without the photocathode - by biasing a 19 stage mesh dynode glass PMT with the photocathode at ~+5V w.r.t. the first dynode, preventing photocathode p.e. from contributing to the dynode signal. [22], [23], [24]. A set of 7 PMTs were mounted on a PC board as shown in Fig 5. The mesh dynode sensor board was tested by placing it behind successive radiation lengths of W-radiators, and exposed behind successive absorbers to 8 and 16 GeV electrons, giving the data of response vs shower depth in Fig 8. *The agreement between data and MC is remarkably good; the prediction for a compensating signal from the dynodes is warranted and should be tested.* From these data we extract that the efficiency for detecting an electron mip from SEe's emitted from the first 5 mesh dynodes is 80%, and that the SEe =1.11 e- per surface.

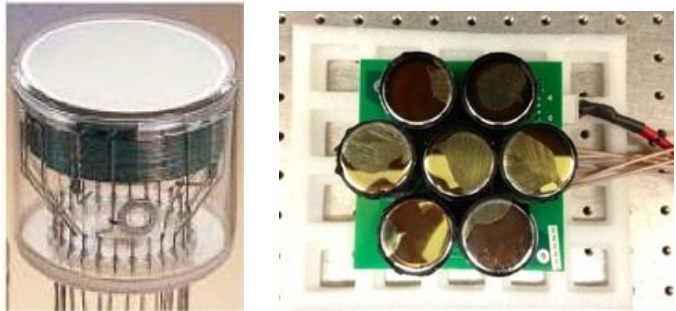

*Figure 5a,b :* Left: 19 stage mesh PMT, and 7 mounted on a base board with the phoocathode shut off to serve as a crude SE calorimeter sensor. The circuitry is selectable to run as normal PMT, or to block photoelectrons from leaving the photocathode by a back potential. During operation, the tube faces were darkened with a flat black paint to help suppress short wavelength Cerenkov multiple bounces from dynode stimulation.

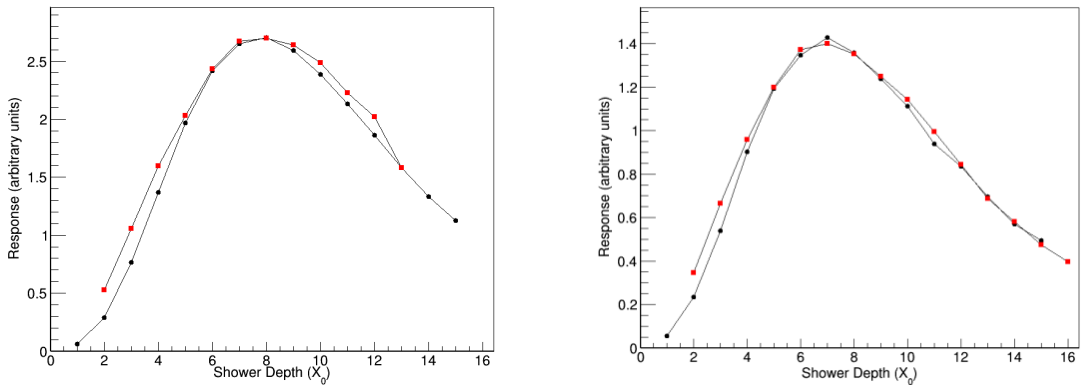

**Figs 5 c,d:** Response of the SE module to 8 (left) and 16 GeV (right) electrons with tungsten absorbers. Data are shown in black and MC simulation results are shown in red.

The measurements shown in Figure 5c,d must be extended to full calorimeters with SE planes as sensors and in hadron showers, and with metal envelope, quartz window PMTs. At present, there are no good tests of PMT as sensors in complete hadron shower calorimeters, either as PMT, or as PMT with the photocathode shut off. Neutron detection by PMT dynodes is has been demonstrated with radiosources and could contribute to compensation. Direct detection of neutrons by PMT is not well-studied. In the few studies done on PMTs, neutrons cause signals dominated by the large (Barns) cross-sections for 1-14 MeV neutrons for $^{10}B(n,\alpha)$, $^{28}Si(n,\alpha)$ and $^{28}Si(n,p)$ leading to luminescence scintillation/glow in borosilicate or soda-lime glass, but otherwise no damage; quartz window metal envelop PMT we are immune from such backgrounds. Signals may result from n's spalling/fissioning absorbers adjacent to PMTs or in PMTs, producing charged particles. We must measure neutron signals from AmBe(~MeV) sources in quartz window+metal PMT, and during neutron irradiation.

### *Photomultipliers as In-Situ Calorimeter Sensors:*
We suggest that photomultiplier and dynodes have a potential as in-situ sensors for energy-flow calorimetry at very high repetition rates and radiation levels, with some potential for compensation and thus better energy resolution than PMT window Cerenkov light alone.

A PMT window has high sensitivity to MIPs due to the Cerenkov light generated in the glass of the window or tube body. In the CMS very forward calorimeter, the photomultipliers produced huge background signals from muons crossing the PMT window and glass vacuum envelope from Cerenkov in the PMT glass and likely also from direct dynode secondary emission electrons (SEe).

We propose to turn that *bug* into a *feature*: Compact PMT with metal envelopes and quartz/sapphire windows are available, able to be manufactured in quantity. An example is a 5x5 cm$^2$ area Hamamatsu PMT with etched metal dynodes and 2x2 up to ~16x16 anodes, only 1.5 cm thick. Similarly mesh-type dynodes can be compact with even better magnetic field resistance. – MAPMTs (Multi-Anode PMT) in present form have the advantage of low inductance and capacitance to ground if care is taken, and, if the dynodes are actively powered (non-resistive base), in principle can operate at 100's MHz with very low hysteresis – rise times of 0.5 ns, and fall times of ~1.5ns, but *more importantly a transit time (i.e the latency – the time the electron cloud traverses the dynodes to the anode) <3ns,* implying ~300 MHz operation is possible. PMT are relatively insensitive to temperature, varying gain to less than ±5% over 0-60 °C range, and with moderate %-gain change with % voltage variation compared with SiPM. If using quartz or sapphire windows, MAPMT are likely insensitive to radiation damage up to GigaRad levels [25]. *With a gain sufficiently low (~5 x10$^4$), the dynode current for 100 MHz of 100 p.e. remains less than 10µA so that dynode heating is not a conditional problem.*

The PMT proposed in-situ for high rate radiation resistant calorimeters must have the following requirements/properties:

- a. Metal Vacuum Envelopes and Quartz or Sapphire windows– no glass glow from radiation.
- b. Radiation Resistant,Thin (≤3mm) Windows of quartz (SiO$_2$), MgF$_2$, or sapphire (Al$_2$O$_3$)
- c. B-Field: Secondary Emission(SE) mesh/metal channeled dynodes, ≥5% max gain @ 1T.
- d. Compact Longitudinal Dimensions: Glass window-to-Anode pins thickness ≤ 1.5cm
- e. Anode Pixel Size: ≥0.3cm x 0.3cm.
- f. Gain: ≥ 5 x 10$^4$ – (Lowest gain necessary to avoid dynode heating at 100 MHz)
- g. Single p.e. risetime ≤1ns; falltime < 1.5ns; slew rate >10 mV/ns into 300 Ω;
- h. Latency/Transit time: T$_{trans}$<3ns enables rate ~ (2 x T$_{trans}$)$^{-1}$ ~ 200 MHz
- i. Photocathode/anode size Area < 3x3mm pixels
- j. QE for Ultra bialkali photocathodes is sufficient at 35%-40%.
- i. Tileable: square or hex shapes

Metal envelope Hamamatsu MAPMT(H12700B) are available with 0.5-0.6ns risetime and 1.5 ns fall time with a latency/transit time ~3ns are shown below. Such tubes can count at 300 MHz with little hysteresis.

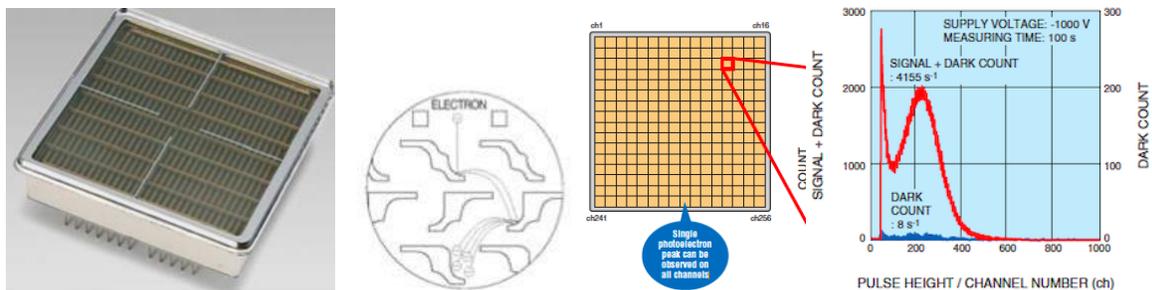

***Fig.6:*** Magnetic field operable, compact 5x5cm x1.4cm end-of-pins-deep MAPMT(H12700B or equiv) using channelized etched dynodes, with 0.55 ns risetime. At right is the superior single p.e. resolution.

The MAPMT is powered with separate adjustable HV power supplies(CAEN 12 channel or equiv), with large buffer capacitors to avoid RC recharging, and a stripline SMA connector to the last dynode and the anode. To test, we propose a pair of Picolas BFS-VRM 03 HP laser diode drivers, capable of up to 400

MHz operation and a pair of 0-300 MHz signal generators to simulate a constant 1-10 p.e. background and a signal; 2 GHz preamps (FEMPTO-HAS-Y series -6 pA/√Hz- or equiv), and a Picoscope series 6000 4 ch, 5 GHz sampling digitizer, or an equivalent or better 4 ch used or rented oscilloscope for DAQ.

**Monte Carlo:** A preliminary MC of a Cerenkov calorimeter (not including direct dynode signals) using towers composed of one hundred 1 cm quartz cubes, with PMT, separated by 3 mm of W, had an energy resolution of 14%/√E in quadrature with a 0.34% constant term from 1-500 GeV (Figs. 7) A point design of a possible front end in Phase III upgrades at CMS is shown in Figs 7. It consists of 5 layers, each of 1000 1cm quartz cubes, 1600 MAPMT, and metal absorber totaling 2 λ and 20 Xo. MC shows it results in a much improved hadronic resolution in the HF forward quartz fiber calorimeter of CMS to 63%/√E with a 7% constant term.

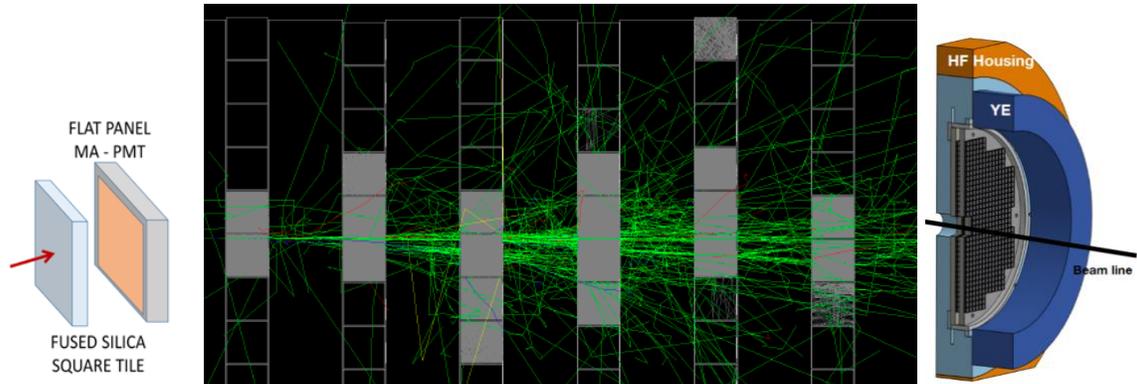

**Figs 7a** Quartz+PMT calorimeter concept.
*Left* is the basic proposed calorimeter component, a rad-hard Cerenkov tile or cube - sensitive to high β/mips coupled to a metal envelope quartz window PMT -if available, rad-hard scintillator tiles could substitute.
*Middle* MC of 100 GeV π into 100 deep quartz cube+PMT+absorber towers
*Right* – Assembly as a forward front end calorimeter - half-cylinder of a 5 layer pre-radiator calorimeter, 3≤η≤5.2, ~10m from an IP, 1m radius, uses1600 5x5cm 16 anode MAPMT(2cm thick)+1cm cube quartz radiators/layer.

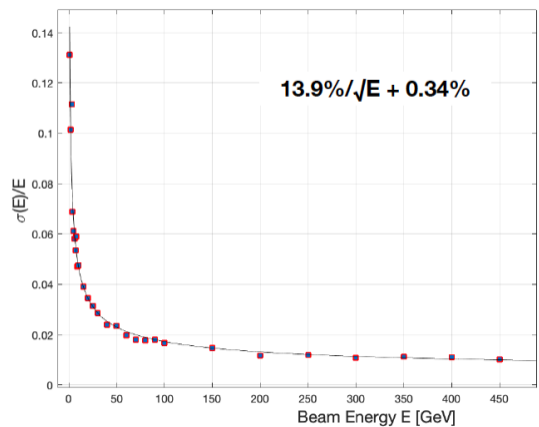

$$13.9\%/\sqrt{E} + 0.34\%$$

**Fig 7b** MC of e-m energy resolution. Direct dynode stack signals are not taken into account in this MC.

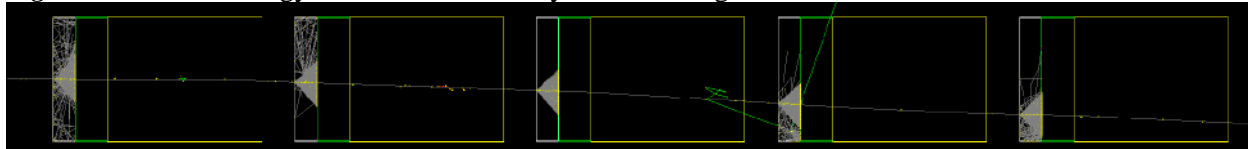

**Fig 7c:** *MC showing Cerenkov light trajectories/cones in Quartz from a muon traversing the 2λ thick preradiator/em particle ID consisting of 5 quartz tiles as mated to MAPMTs, and iron absorber.*

The most reliably radiation resistant (few GRads) and very fast optical signals for calorimetry are from Cerenkov radiators – tiles of quartz, synthetic sapphire ($Al_2O_3$), silica aerogels, and $MgF_2$. In the future, in-situ PMTs coupled to scintillators would have higher signals, provided the scintillator has radiation resistance and the optical signals are fast. At present, the most promising such fast, high-rate tiles are ZnO:Ga tiles - 0.7ns decay time – the largest number of photons per ns of any known scintillator, 20% of the integrated output of LYSO(42 ns decay time), or quartz tiles loaded with nanoparticles such as nitrogen-defect nanodiamond or other inorganic nano-scintillators. There are no *convenient* hydrogenous scintillators to enhance the hadron-neutron signal.

### Prerequisites for a very high rate preradiator and $\eta > 3$:

*1. PMT with quartz or synthetic sapphire windows, metal envelope and conpact dynode stack.*
*2. Individual dynode powering with high density cabling - Rate, Hysteresis, $\sigma_T$, Dynamic Range*
*3. Measured PMT neutron and low energy particle (alpha, proton) Response:*
*4. Radiation damage of operating quartz window metal envelope PMTs to $\gamma$ and $n$:*

### Summary/Objectives:

In the Energy Frontier we suggest developing high rate ($\geq 100$ MHz) finely segmented forward calorimetry preradiators with time resolution <50 ps which will survive the first ~1-2 $\lambda$ of incident high radiation doses, protecting forward calorimeters ($3 < \eta < 6$; less than $5°$ to the beam) behind them from radiation damage, with high granularity, high rate capability and ~40ps time resolution (4D-calorimetry) providing lepton/photon ID and measurement. In the Intensity Frontier beam particle selection, such as tagged neutrino and kaon beams, and lepton violation experiments with muons require very high rates. Cosmic Frontiers requiring low power, non-cooled calorimetry or optical detection that can keep track of particles or photons arriving at 100's of MHz, and survivable for years in space radiation may also benefit. The basic research is to use compact channelized PMTs with quartz or other radiation resistant windows with metal envelopes as an in-situ sensor, directly coupled to Cerenkov (or radiation-resistant scintillator) tiles, utilizing the dynode signals as a potentially compensating 2nd signal, and with no active electronics. If successful, directions include proposals for high SE yield mesh dynode activator materials such as GaP or B-doped diamond films with 25 SEe at 300 eV electron energies, and possibly for compact low cost tile SE sensors – no photocathode - far easier to fabricate than PMTs: all metal final assembly in air, brazed seals; bake-out 900°C; pump out with tip-off - vacuum 100x higher than PMTs. Such sensors have many applications beyond HEP, in research, medicine, industry and defense.